\documentclass[cits]{PoS}

\usepackage[utf8]{inputenc}

\usepackage{xcolor}

\usepackage{mathtools}

\usepackage{booktabs}

\usepackage{microtype}

\usepackage{cite}

\usepackage{tikz}
\usetikzlibrary{calc}

\definecolor{fu-red}{RGB}{204, 0, 0}

\title{Non-perturbative renormalization of $\mathrm{O}(a)$ improved tensor currents}

\ShortTitle{Non-perturbative renormalization of $\mathrm{O}(a)$ improved tensor currents}

\author{
\begin{minipage}[b]{0.4\linewidth}
\includegraphics[height=2.5\baselineskip]{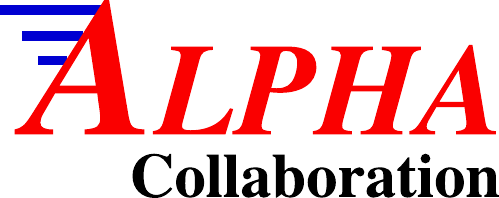}
\end{minipage}
\hfill \\
\hfill\parbox{30.5mm}{\vspace{-2.cm}\raggedleft\footnotesize\it
	CERN-TH-2019-146\\ 
	MS-TP-19-21\\
	IFT-UAM/CSIC-19-123
}
}

\vspace{1cm}

\author{Leonardo~Chimirri$^a$, Patrick~Fritzsch$^b$, Jochen~Heitger$^c$,  \speaker{Fabian~Joswig}$^{\,c}$, Marco~Panero$^{de}$, Carlos~Pena$^{f}$ and David~Preti$^{e}$\\
\llap{$^a$}John von Neumann Institute for Computing (NIC), DESY\\
Platanenallee 6, 15738 Zeuthen, Germany\\
\llap{$^b$}Theoretical Physics Department, CERN\\
1211 Geneva 23, Switzerland\\
\llap{$^c$}Institut f\"ur Theoretische Physik, Westf\"alische Wilhelms-Universit\"at M\"unster\\
Wilhelm-Klemm-Straße 9, 48149 M\"unster, Germany\\
\llap{$^d$}Department of Physics, University of Turin\\
Via Pietro Giuria 1, 10125 Turin, Italy\\
\llap{$^e$}INFN, Turin\\
Via Pietro Giuria 1, 10125 Turin, Italy\\
\llap{$^f$} Instituto de F\'{\i}sica Te\'orica UAM/CSIC and Departamento de F\'{\i}sica Te\'orica\\
Calle Nicol\'as Cabrera 13-15, Cantoblanco, 28049 Madrid, Spain\\

E-mail: \email{fabian.joswig@wwu.de}}

\abstract{We present our progress in the non-perturbative $\mathrm{O}(a)$ improvement and renormalization of tensor currents in three-flavor lattice QCD with Wilson-clover fermions and tree-level Symanzik improved gauge action. The mass-independent ${\rm O}(a)$ improvement factor of tensor currents is determined via a Ward identity approach, and their renormalization group running is calculated via recursive finite-size scaling techniques, both implemented within the Schr\"odinger functional framework. We also address the matching factor between bare and renormalization group invariant currents for a range of lattice spacings $<0.1\,$fm, relevant for phenomenological large-volume lattice QCD applications.
}

\FullConference{37th International Symposium on Lattice Field Theory - Lattice2019\\
		16-22 June 2019\\
		Wuhan, China}

\begin{document}

\section{Introduction}

The study of weak matrix elements is relevant for some very interesting processes to test the Standard Model (SM), which may have the potential to reveal New Physics: these include rare decays of heavy mesons, observables related to $\beta$-decays of the neutron, and effects that could be important for direct detection of dark matter. As such processes occur in color-singlet hadrons, an environment dominated by non-perturbative effects of the strong interaction, their systematic study from the first principles of quantum chromodynamics (QCD) is one of the main tasks that can be successfully approached through lattice calculations.

In the SM, these decays can be treated via an effective Hamiltonian that encodes weak dynamics in terms of effective quark field interactions at hadronic energies. Here we focus on one such possible interaction terms, namely, flavor non-singlet currents with a tensor Dirac structure. They have the form
\begin{align}
\label{tensor_current_continuum_definition}
T^a_{\mu\nu}=i\bar{\psi}(x)\sigma_{\mu\nu} T^a \psi(x)\,,
\end{align}
where $T^a$ is a generator of the $\mathrm{SU}(N_\mathrm{f})$ acting on the flavor indices, while $\sigma_{\mu\nu}=(i/2)[\gamma_{\mu},\,\gamma_{\nu}]$ acts on spinor indices. Partial-current-conservation laws imply that flavor non-singlet vector and axial currents do not require ultraviolet renormalization such that the scale dependence of the quark mass stems directly from the anomalous dimensions of the scalar and pseudo-scalar densities. Tensor-like currents of the form~(\ref{tensor_current_continuum_definition}), however, are not constrained by such laws, and require an independent scale-dependent renormalization. Perturbative calculations of the anomalous dimension associated with these currents have been carried out in the modified minimal-subtraction~\cite{Gracey:2000am} and in the regularization-invariant momentum-subtraction schemes~\cite{Almeida:2010ns} in the continuum, as well as in lattice schemes~\cite{Skouroupathis:2008mf}.

In an effort within the ALPHA Collaboration, we study the renormalization of these currents non-perturbatively, on a set of lattice configurations with $N_f=2+1$ dynamical Wilson-clover quarks and either plaquette or tree-level Symanzik-improved gauge actions, depending on convenience. This extends the analysis previously carried out in ref.~\cite{Pena:2017hct} on configurations with $N_f=0$ and $N_f=2$ dynamical flavors. In addition, we determine the coefficient of the $\mathrm{O}(a)$ improvement term of the local tensor current operator built from Wilson quarks, to achieve better convergence to the continuum limit. A closely related work, analyzing the quark-mass renormalization and running, is reported in ref.~\cite{Campos:2018ahf}, whereas studies addressing the vector and axial current cases include refs.~\cite{Bulava:2015bxa, Bulava:2016ktf, DallaBrida:2018tpn,Heitger:2017njs, Fritzsch:2018zym, Gerardin:2018kpy, Korcyl:2016ugy}.

\section{Calculation set-up in the Schr\"odinger functional formalism}

We work in the Schr\"odinger functional formalism, which provides a powerful regularization-independent renormalization scheme~\cite{Luscher:1992an, Luscher:1993gh, Sint:1993un}. It is based on the idea of formulating QCD in a system of finite linear size $L$, with periodic boundary conditions (b.c.) in the spatial directions and fixed b.c.\ in the time direction of extent $T$. On the one hand, this allows one to define a non-perturbative scheme for the QCD coupling, and to study its evolution with the scale defined by the inverse of the system size. On the other hand, it also endows the theory with an explicit infrared cut-off, streamlining the operator improvement and mass-independent renormalization procedures.

From the Schrödinger functional boundary fields $\zeta$, $\bar{\zeta}$, $\zeta^\prime$ and $\bar{\zeta}^\prime$ (see \cite{Sint:1993un}), we define boundary-to-bulk correlation functions,
\begin{align}
k_\mathrm{T}(x_0)=-\frac{a^6}{6}\sum_{\mathbf{u},\mathbf{v}} \big\langle T_{0k}(x_0)\bar{\zeta}(\mathbf{u})\gamma_k\zeta(\mathbf{v})  \big\rangle\,, \quad k_\mathrm{V}(x_0)=-\frac{a^6}{6}\sum_{\mathbf{u},\mathbf{v}} \big\langle V_{k}(x_0)\bar{\zeta}(\mathbf{u})\gamma_k\zeta(\mathbf{v}) \big\rangle\,, \label{eq:k_correlationfunctions}
\end{align}
as well as boundary-to-boundary ones like
\begin{align}
k_1=-\frac{a^{12}}{6L^6}\sum_{\mathbf{u},\mathbf{v},\mathbf{u}^\prime,\mathbf{v}^\prime}\big\langle \bar{\zeta}^\prime(\mathbf{u}^\prime)\gamma_k\zeta^\prime(\mathbf{v}^\prime) \, \bar{\zeta}(\mathbf{u})\gamma_k\zeta(\mathbf{v}) \big\rangle\,.
\end{align}

\section{Non-perturbative improvement of the tensor current}

\begin{figure}[htb]
	\centering
	\graphicspath{{figures/}}
	\resizebox{0.5\textwidth}{!}{\input{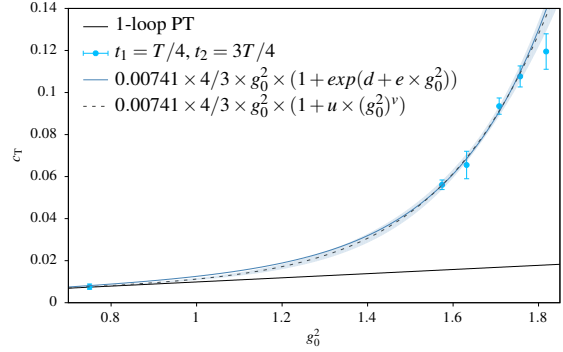}}
	\caption{Approach of the coefficient $c_\mathrm{T}$, evaluated non-perturbatively, as discussed in the text, towards its one-loop approximation~\cite{Taniguchi:1998pf} (solid black line). The solid  blue and the dashed black curves are obtained from two-parameter fits to the functional forms displayed in the legend.}
	\label{fig:c_T_Q0_final}
\end{figure}

The lattice formulation with Wilson fermions has $\mathrm{O}(a)$ discretization errors, which affect both the action and local operators. These lattice artifacts can be canceled by appropriate counterterms, allowing one to achieve $\mathrm{O}(a^2)$ scaling towards the continuum limit~\cite{Luscher:1996sc}. Restricting to the zero-momentum channel, only the ``electric'' components of the tensor current require this improvement; we then define the improved version of the tensor current as
\begin{align}
\big(T_{0k}^a\big)^\mathrm{I}=T_{0k}^a+ac_\mathrm{T}(g_0^2)\tilde{\partial}_0 V_k^a\,,\quad \big(T_{ij}^a\big)^\mathrm{I}=T_{ij}^a\,. \label{eq:improvedtensorcurrent}
\end{align}
As the $c_\mathrm{T}$ coefficient is known perturbatively only at one-loop~\cite{Taniguchi:1998pf}, we compute it non-perturbatively, using the following chiral Ward identity, inspired by the discussion on chiral Ward identities in~\cite{Bhattacharya:1999uq}:
\begin{eqnarray}
\label{chiral_Ward_identity_for_cT}
&& \hspace{-1cm} \epsilon_{0kij}\bigg( \int \mathrm{d}^3\mathbf{x}\Big\langle \big[A_0^a(t_2,\mathbf{x})-A_0^a(t_1,\mathbf{x})\big]{T}_{ij}^b(y)\mathcal{O}_\mathrm{ext}\Big\rangle 
-2m \int \mathrm{d}^3\mathbf{x} \int_{t_1}^{t_2}\mathrm{d}x_0\,\Big\langle P^a(x_0,\mathbf{x}){T}_{ij}^b(y)\mathcal{O}_\mathrm{ext}\Big\rangle\bigg) \nonumber \\
&& \qquad =2d^{abc}\,\Big\langle T_{0k}^c(y)\mathcal{O}_\mathrm{ext}\Big\rangle\,. 
\end{eqnarray}
The Ward identity is evaluated non-perturbatively on gauge configurations with Schrödinger functional b.c.\ also employed in previous works \cite{Bulava:2015bxa,deDivitiis:2019xla}.
\begin{figure}[b]
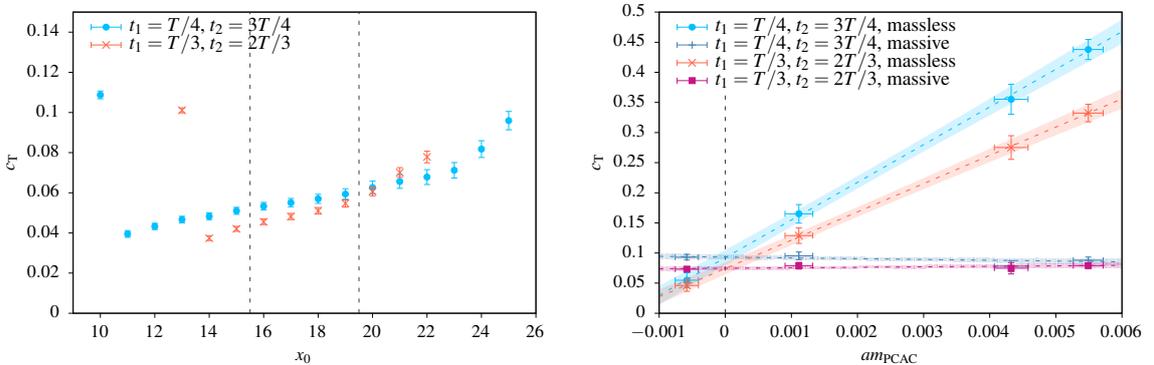

	\centering
	\begin{minipage}{.49\textwidth}
		\centering
		\graphicspath{{figures/}}
		\resizebox{\textwidth}{!}{\input{figures/c_T_plateau_D1k4.tex}}
	\end{minipage}
	\hfill
	\begin{minipage}{0.49\textwidth}
		\centering
		\graphicspath{{figures/}}
		\resizebox{\textwidth}{!}{\input{figures/B_0_m_chiral.tex}}
	\end{minipage}
	\caption{Left: $c_\mathrm{T}$ values extracted using eq.~(\ref{chiral_Ward_identity_for_cT}), from configurations on a $(35 \times 24^3)a^4$ lattice with $\beta = 3.81$ and $am_{\mbox{\tiny{PCAC}}} = -0.00003(7)$. Right: $c_\mathrm{T}$ values at $\beta = 3.512$ for different quark masses.}
	\label{fig:double_fig}
\end{figure}
The ensembles lie on a line of constant physics with a spatial extent of $L\approx 1.2\,$fm, which ensures that all ambiguities (in imposing a specific improvement condition) proportional to the lattice spacing vanish smoothly towards the continuum limit.
The left panel of figure~\ref{fig:double_fig} shows the $c_\mathrm{T}$ values extracted in this way, on a lattice of hypervolume $(35 \times 24^3)a^4$ at $\beta = 3.81$ with $am_{\mbox{\tiny{PCAC}}} = -0.00003(7)$. The plot on the right reports the $c_\mathrm{T}$ values obtained at $\beta = 3.512$ for different quark masses, and shows the robustness of the values obtained at zero quark mass in the Schr\"odinger functional formalism.
In figure~\ref{fig:c_T_Q0_final}, results for $c_\mathrm{T}$ in a range of bare couplings relevant for large volume simulations are shown. To check that a smooth connection to the perturbative one-loop prediction is made, an additional simulation in the perturbative regime ($g_0^2=0.75$) has been conducted, which supports the expected asymptotics of the non-perturbative improvement coefficient. Some more technical details, such as the choice of the external operator $\mathcal{O}_\mathrm{ext}$, as well as the operator placements $t_1$ and $t_2$ in figure~\ref{fig:double_fig}, will be explained in \cite{chimirri:2019}.

\section{Renormalization}
Next, we discuss the renormalization of the tensor current. Our strategy follows the standard
non-perturbative renormalization and running setup by the ALPHA Collaboration, in particular
in the context of $N_{\rm\scriptscriptstyle f}=3$ QCD.
We define the renormalization group invariant (RGI) tensor current
\begin{gather}
\hat{T}_{\mu\nu} = \overline{T}_{\mu\nu}(\mu) \left[\frac{\overline{g}^2(\mu)}{4\pi}\right]^{-\gamma_0/2b_0}
\exp\left\{
-\int_0^{\overline{g}(\mu)}{\rm d}g\left[\frac{\gamma(g)}{\beta(g)}-\frac{\gamma_0}{b_0g}\right]
\right\}\,,
\end{gather}
where $\overline{T}_{\mu\nu}(\mu)$ is the renormalized tensor current in the continuum,
$\overline{g}(\mu)$ is some renormalized coupling, $\beta$ and $\gamma$ are the $\beta$-function and the tensor anomalous dimension,
respectively, and $b_0,\gamma_0$ their leading perturbative coefficients.
We define the multiplicative renormalization factor $Z_\mathrm{T}(g_0^2,a/L)$ from the mass-independent, finite-volume renormalization condition
\begin{align}
Z_\mathrm{T}(g_0^2,a/L)\cdot \frac{k_\mathrm{T}^\mathrm{I}(L/2)}{\sqrt{k_1}}=\frac{k_\mathrm{T}^\mathrm{I}(L/2)}{\sqrt{k_1}}\Bigg|_{\mbox{\tiny{tree level}}}\,,
\end{align}
where $k_\mathrm{T}^\mathrm{I}(x_0)$ is defined according to eqs.\ (\ref{eq:k_correlationfunctions}) and (\ref{eq:improvedtensorcurrent}).
This serves two purposes: we can renormalize the tensor current at any given scale $\mu=1/L$, through
\begin{gather}
\overline{T}_{0k}(\mu) = \lim_{a\to 0}Z_\mathrm{T}(g_0^2,a/L)T_{0k}(g_0^2)\,,
\end{gather}
where $T_{0k}(g_0^2)$ stands for the insertion of the tensor current in a bare correlation function computed at bare coupling $g_0^2$, and trace the renormalization group evolution of the current
by introducing the step scaling function (SSF)
\begin{gather}
\sigma_{\rm\scriptscriptstyle T}(u) \equiv \lim_{a\to 0}\Sigma_{\rm\scriptscriptstyle T}(u,a/L)
\equiv \lim_{a\to 0} \left.\frac{Z_\mathrm{T}(g_0^2,a/(2L))}{Z_\mathrm{T}(g_0^2,a/L)}\right|_{u=\overline{g}^2(1/L)}\,.
\end{gather}
By computing $\Sigma_{\rm\scriptscriptstyle T}(u,a/L)$ at several values of $u$ and $a/L$ it is possible to
obtain $\sigma_{\rm\scriptscriptstyle T}(u)$, and hence $\gamma$, non-perturbatively for a wide range of scales,
using convenient finite-volume schemes. Our final expression for the RGI current will read
\begin{gather}
\label{eq:master}
\hat{T}_{\mu\nu} =
\underbrace{\frac{\hat{T}_{\mu\nu}}{\overline{T}_{\mu\nu}(\mu_{\rm\scriptscriptstyle pt})}}_\text{PT}\,
\underbrace{\frac{\overline{T}_{\mu\nu}(\mu_{\rm\scriptscriptstyle pt})}{\overline{T}_{\mu\nu}(\mu_0/2)}}_\text{SF}\,
\underbrace{\frac{\overline{T}_{\mu\nu}(\mu_0/2)}{\overline{T}_{\mu\nu}(\mu_{\rm\scriptscriptstyle had})}}_\text{GF}\,
\overline{T}_{\mu\nu}(\mu_{\rm\scriptscriptstyle had})\,,
\end{gather}
where $\mu_{\rm\scriptscriptstyle had}$ is some low-energy scale $\sim\Lambda_{\rm\scriptscriptstyle QCD}$,
$\mu_{\rm\scriptscriptstyle pt}$ is some high-energy scale $\sim M_W$ where perturbation theory is safe (NLO is available in our case),
$\mu_0$ is an intermediate scale $\sim 4~{\rm GeV}$,
and the factors labeled ``GF'' and ``SF'' are computed using gradient flow and SF non-perturbative couplings,
respectively
(see~\cite{Campos:2018ahf} for a detailed explanation, full reference list, and any unexplained notation).
The key points in the whole setup are that each of these factors, except for the first one,
can be computed non-perturbatively and taken to the continuum limit with fully controlled systematics,
and that the connection to the RGI allows to match the result to any other renormalization scheme convenient for phenomenology.

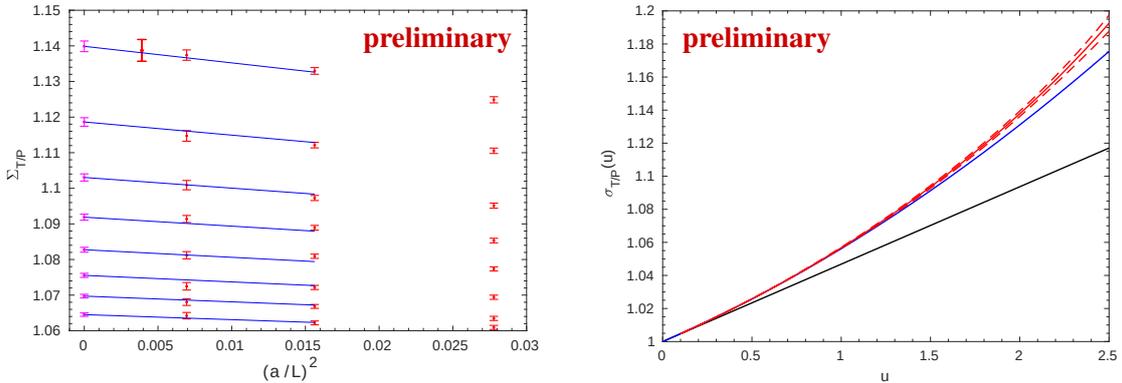
\begin{figure}[b]
	\centering
	\begin{minipage}{.49\textwidth}
		\centering
		\includegraphics[height=5cm]{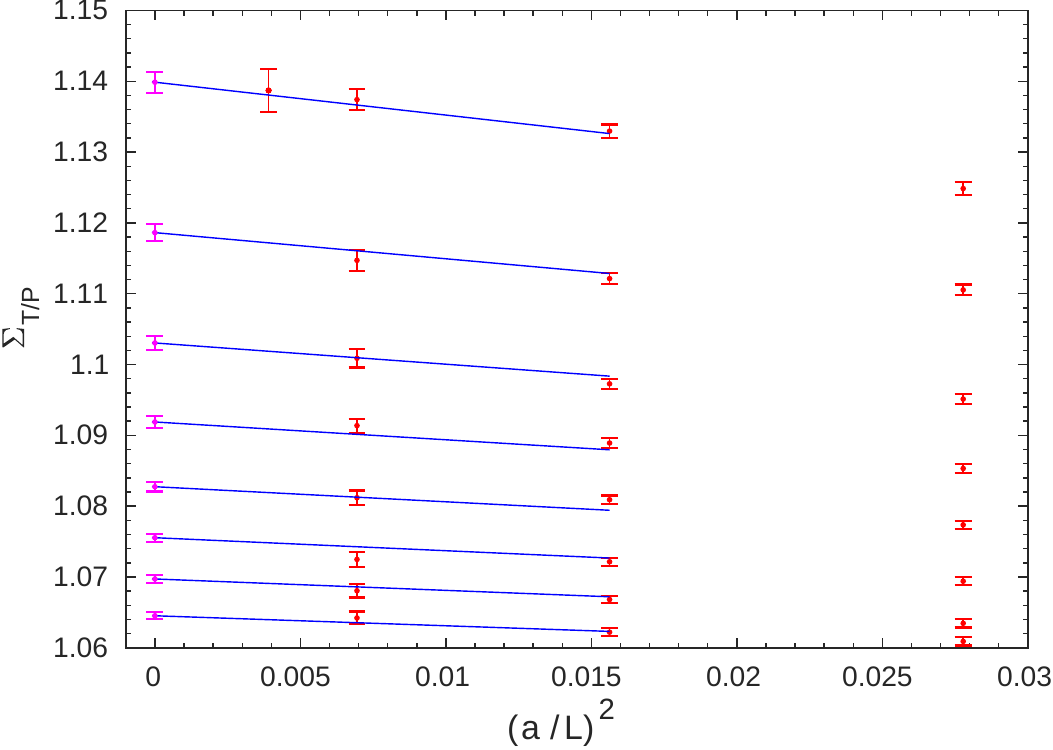}
		\begin{tikzpicture}[remember picture, overlay]
		\node[anchor=south east,inner sep=0pt] at ($(current page.north east)-(8.7cm,17.2cm)$) {
			\textbf{\color{fu-red}preliminary}
		};
		\end{tikzpicture}
	\end{minipage}
	\hfill
	\begin{minipage}{0.49\textwidth}
		\centering
		\includegraphics[height=5cm]{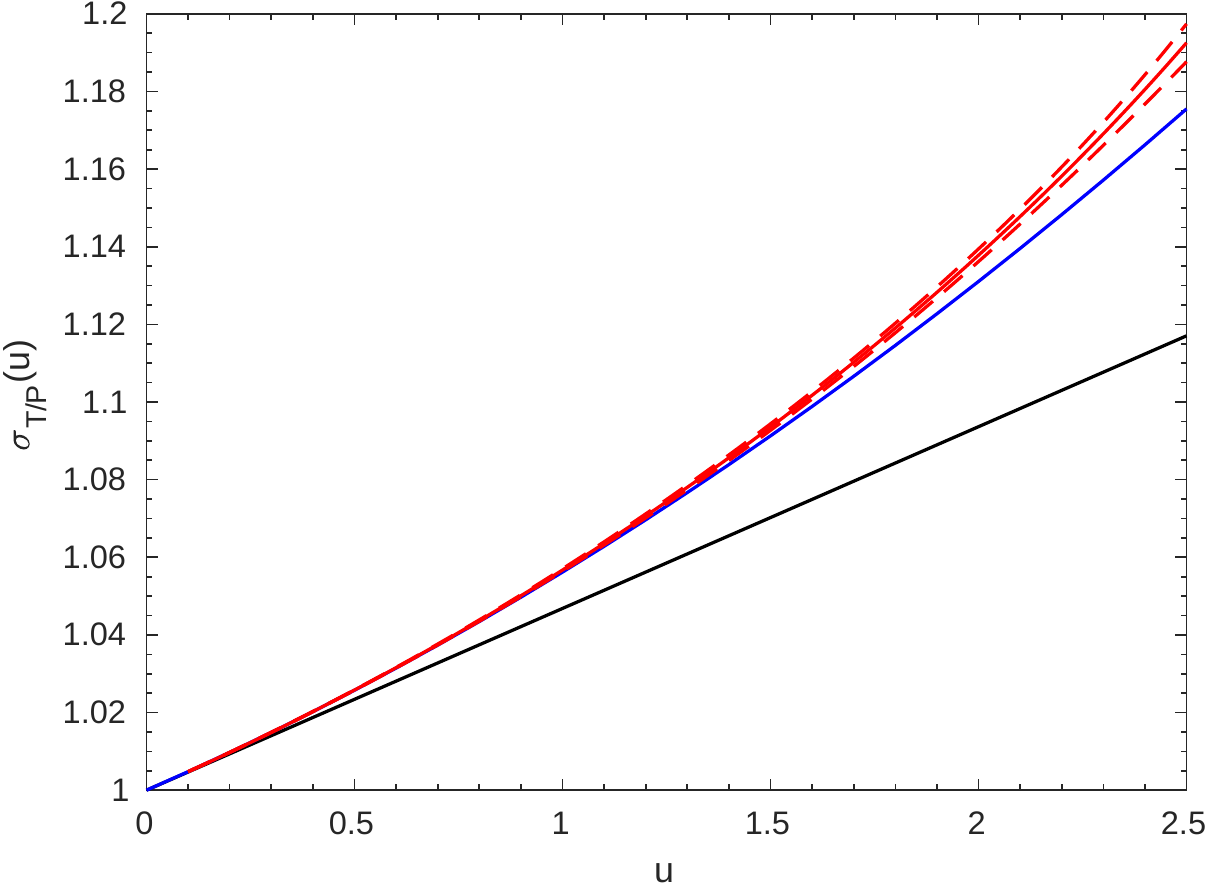}
		\begin{tikzpicture}[remember picture, overlay]  
		\node[anchor=south east,inner sep=0pt] at ($(current page.north east)-(4.5cm,17.2cm)$) {
			\textbf{\color{fu-red}preliminary}
		};
		\end{tikzpicture}
	\end{minipage}
	\caption{Left: Examples of results for the lattice SSF built from the $Z_\mathrm{T}/Z_\mathrm{P}$ ratio, for different values of the squared SF coupling, denoted by $u$. These results are extrapolated linearly in $(a/L)^2$ to the continuum SSF $\sigma_\mathrm{T/P}$. Right: Numerical results (red) compared with the one- (black) and two-loop (blue) PT predictions.}
	\label{fig:step-scaling_functions}
\end{figure}

\begin{figure}[htb]
	\centering
	\includegraphics[width=0.5\textwidth]{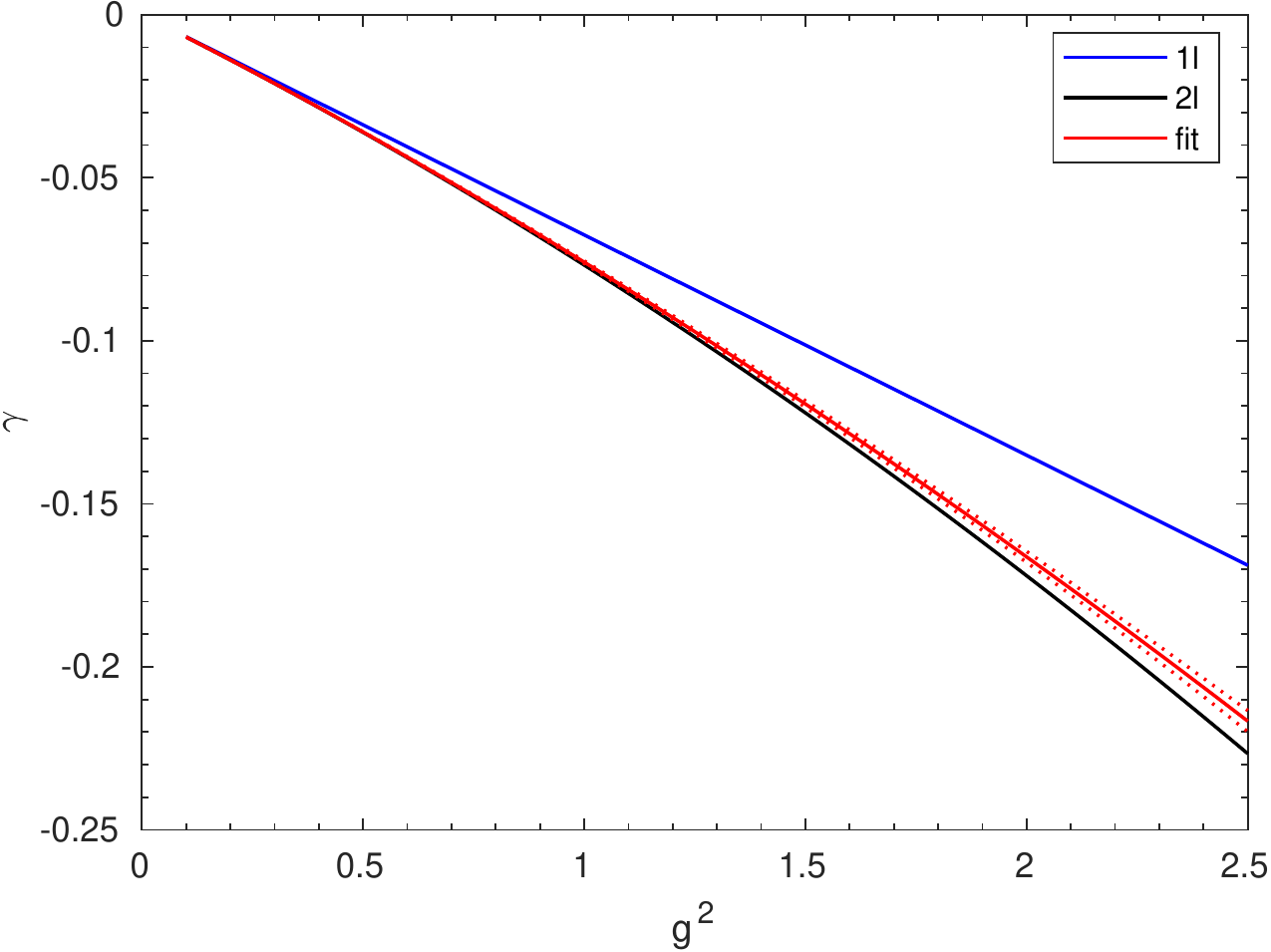}
	\begin{tikzpicture}[remember picture, overlay]  
	\node[anchor=south east,inner sep=0pt] at ($(current page.north east)-(8.8cm,5.5cm)$) {
		\textbf{\color{fu-red}preliminary}
	};
	\end{tikzpicture}
	\caption{The anomalous dimension derived from the $Z_\mathrm{T}/Z_\mathrm{P}$ ratio, denoted by $\gamma$, against the squared coupling in the SF scheme. The fit to our results, represented by the red curve, is compared with the one- and two-loop (``1l'' and ``2l'') perturbative predictions.}
	\label{fig:anomalous}
\end{figure}
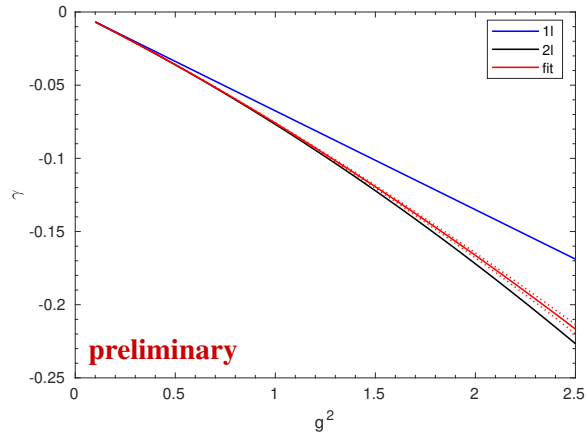

We now present first results for the renormalization group evolution in the intermediate-coupling regime, i.e., between the scales $\mu_0/2$ and $\mu_{\rm\scriptscriptstyle pt}$;
results for the evolution between $\mu_{\rm\scriptscriptstyle had}$ and $\mu_0/2$ are underway~\cite{chimirri:2019}
The continuum-limit extrapolation for $\sigma_{\rm\scriptscriptstyle T}(u)$ is illustrated in figure~\ref{fig:step-scaling_functions}, which also highlights a comparison with the one- and two-loop perturbative predictions for the continuum SSF.
In figure~\ref{fig:anomalous} we present our numerical results for the anomalous dimension extracted from the $Z_\mathrm{T}/Z_\mathrm{P}$ ratio, comparing them with the perturbative one- and two-loop predictions.
\section{Conclusions and outlook}
In this contribution, we presented the preliminary status of a non-perturbative study of bilinear flavor non-singlet tensor currents that are of particular relevance for phenomenology. We have already completed the calculation of the improvement coefficient for the tensor current along a line of constant physics, as well as its renormalization at intermediate energy scales. As a byproduct, we also obtained the anomalous dimension $\gamma$ associated with the tensor current (the only non-vanishing anomalous dimension for these currents that is independent from the one of the quark mass). We are currently extending the renormalization study to the low-energy domain in the GF scheme: this analysis is at an advanced stage, and, by combining its final results with those at higher momentum scales, we will be able to obtain the complete evolution of the renormalization factor for the tensor current from the hadronic scale to the high-energy limit along eq.~(\ref{eq:master}) \cite{chimirri:2019}. Another study of the tensor current renormalization constant in the $\text{RI}^\prime$-MOM scheme with the same set of actions was pursued in \cite{Harris:2019bih} and can be used for comparison to our final results.
\\
\\
{\footnotesize
\textbf{Acknowledgments:} This work is supported by the Deutsche Forschungsgemeinschaft (DFG) through the Research Training Group \textit{``GRK 2149: Strong and Weak Interactions -- from Hadrons to Dark Matter''} (J. H. and F. J.). We acknowledge the computer resources provided by the \textit{Zentrum f\"ur Informationsverarbeitung} of the University of M\"unster (PALMA II) and thank its staff for support. This project has received funding from the European Union’s Horizon 2020 research and innovation programme under the Marie Skłodowska-Curie grant agreement No. 813942 (LC). C.P. acknowledges support from the EU H2020-MSCA-ITN-2018-813942 (EuroPLEx), Spanish MICINN and MINECO grants FPA2015-68541-P (MINECO/FEDER) and PGC2018-094857-B-I00, and the Spanish Agencia Estatal de Investigaci\'on through the grant ”IFT Centro de Excelencia Severo Ochoa SEV-2016-0597”.
}
{\small
\bibliographystyle{JHEP}
\bibliography{bibliography}
}

\end{document}